\documentclass[useAMS,usenatbibi,usegraphicx]{mn2e}
\usepackage{psfig}

\newcommand\beq{\begin{equation}}
\newcommand\eeq{\end{equation}}

\def\msun{\,{\rm M_\odot}}
\def\etal{{et al.\ }}
\def\gsim{ \lower .75ex \hbox{$\sim$} \llap{\raise .27ex \hbox{$>$}} }
\def\lsim{ \lower .75ex\hbox{$\sim$} \llap{\raise .27ex \hbox{$<$}} }

\title[Gravitational waves from black hole seeds]
{The imprint of massive black hole formation models on the {\it LISA} data stream}

\author[A. Sesana et al.]{Alberto Sesana$^{1}$, Marta Volonteri$^{2,3}$ \& Francesco Haardt$^{1}$\\
$^{1}$Dipartimento di Fisica \& Matematica, Universit\'a dell'Insubria,
via Valleggio 11, 22100 Como, Italy\\
$^{2}$Department of Physics and Astronomy, Northwestern University, 2145 Sheridan Avenue, Evanston, IL, USA\\
$^{3}$Department of Astronomy, University of Michigan, 500 Church Street, Ann Arbor, MI, USA}

\begin{document}

\date{Received ---}

\maketitle

\begin{abstract}
The formation, merging, and accretion history of massive black holes along the 
hierarchical build--up of cosmic structures leaves a unique imprint on the background 
of gravitational waves at mHz frequencies. We study here, by means of dedicated 
simulations of black hole build--up, the possibility of constraining 
different models of black hole cosmic evolution using future gravitational wave 
space--borne missions, such as {\it LISA}. We consider two main scenarios for black hole 
formation, namely, one where seeds are light ($\simeq 10^2 \msun$, remnant of 
Population III stars),  and one where seeds are heavy ($\gsim 10^4 \msun$, 
direct collapse).   In all the models we have investigated, massive black hole 
binary coalescences do not produce a stochastic GW background, but rather, a 
set of individual resolved events.  Detection of several hundreds merging events 
in a 3 year {\it LISA} mission will be the sign of a heavy seed scenario with  
efficient formation of black hole seeds in a large fraction of high redshift halos. 
At the other extreme, a low event rate, about a few tens in 3 years,  is peculiar 
of scenarios where either the seeds are light, and many coalescences do not fall 
into the {\it LISA} band, or seeds are massive, but rare.  In this case a decisive 
diagnostic is provided by the shape of the mass distribution of detected events. 
Light binaries ($m<10^4\msun$) are predicted in a  fairly large number in Population 
III  remnant models, but  are totally absent in direct collapse models.  Finally, a 
further, helpful diagnostic of black hole formation models lies in the distribution 
of the mass ratios in binary  coalescences. While heavy seed models predict that most 
of the detected events involve equal mass binaries, in the case of light seeds, 
mass ratios are equally distributed in the range $0.1-1$ .

\end{abstract}

\section{Introduction}
Massive black hole (MBH) binaries (MBHBs) are among the primary candidate 
sources of gravitational waves (GWs) at mHz frequencies (see, e.g., 
Haehnelt 1994; Jaffe \& Backer 2003; Wyithe \& Loeb 2003, Sesana \etal 2004, 
Sesana \etal 2005), the range probed by the space-based {\it Laser 
Interferometer Space Antenna} ({\it LISA}, Bender \etal 1994). 
Today, MBHs are ubiquitous in the 
nuclei of nearby galaxies (see, e.g., Magorrian et al. 1998). If MBHs were 
also common in the past (as implied by the notion that many distant galaxies 
harbour active nuclei for a short period of their life), and if their host 
galaxies experience multiple mergers during their lifetime, as dictated by 
popular cold dark matter (CDM) hierarchical cosmologies, then MBHBs inevitably formed  
in large numbers during cosmic history. MBHBs that 
are able to coalesce in less than a Hubble time (as defined at the epoch
of their formation) give origin to the 
loudest GW signals in the Universe.
Provided MBHBs do not ``stall'', their GW driven inspiral will then 
follow the merger of galaxies and protogalactic structures at high redshifts. 
A low--frequency detector like {\it LISA} will be sensitive to GWs from coalescing binaries 
with total masses in the range $10^3-10^6\,\msun$ out to $z\sim 5-10$ (Hughes 
2002). 
Two outstanding questions are then how far up in the dark halo
merger hierarchy do MBHs form, and whether stellar and/or gas dynamical processes 
can efficiently drive wide MBHBs to the GW emission stage.

Today we know that MBHs must have been formed early in the history of the Universe. 
Indeed, the luminous $z\approx 6$ quasars discovered in the Sloan Digital Sky 
Survey (Fan \etal 2001) imply that black holes  more massive than a few 
billion solar masses were already assembled when the universe was less 
than a billion years old. Several 
scenarios have been proposed for the seed MBH formation:
seeds of $m_{\rm seed}\sim $few$\times100\msun$ can 
form as remnants of metal free (PopIII) stars 
at redshift $\gsim20$ (Volonteri, Haardt \& Madau 2003, hereinafter VHM),  
while intermediate--mass seeds ($m_{\rm seed}\sim10^5\msun$) 
can be the endproduct of the dynamical instabilities arising in  
massive gaseous protogalactic disks in the redshift range $10\lsim z \lsim15$ 
(Koushiappas, Bullock \& Dekel 2004, hereinafter KBD; Begelman, Volonteri
\& Rees 2006, hereinafter BVR, Lodato \& Natarajan 2006).
All these models have proved successful in reproducing 
the AGN optical luminosity function in a large redshift range ($1\la z\la6$), 
but result in different coalescence rates of MBHBs, and hence in different GW backgrounds. 

In this paper we use the computational tools developed in
Sesana \etal 2005, to characterize the expected GW signal from 
inspiraling MBHBs in the different seed formation scenarios.
Our aim is to understand the {\it LISA} capability to  
place constraints on MBH formation scenarios prior to 
the reionization epoch, looking for reliable diagnostics 
to discriminate between the different models.  

The paper is organized as follows. In \S~2 we describe
the different proposed seed formation scenarios.
In \S~3 we summarize the basic of the detection of 
GW from MBHBs. In \S~4 we compare the {\it LISA} detection rate and the 
properties of the detected MBHB population arising from the different
seed formation scenarios. Finally, we summarize our main 
results in \S~5. Unless otherwise stated, all results shown below refer to 
the currently favored $\Lambda$CDM world model with $\Omega_M=0.3$, 
$\Omega_\Lambda=0.7$, $h=0.7$, $\Omega_b=0.045$, $\sigma_8=0.93$, and $n=1$.

\section{Models of black hole formation}

In our hierarchical framework MBHs grow starting from pregalactic 
seed MBHs formed at early times.
The merger process would inevitably form a large number 
of MBHBs during cosmic history.
Nuclear activity is triggered by halo mergers: in each major merger 
the more massive hole accretes gas until its mass scales 
with the fifth power of the circular velocity of the host halo,  
normalized to reproduce the observed local correlation 
between MBH mass and velocity dispersion ($m_{\rm BH}-\sigma_*$ relation).  
Gas accretion onto the MBHs is assumed to occur at a fraction of the Eddington rate, as empirically shown in simulations of AGN feedback in a merger driven scenario. The scaling we adopt is based on the fitting formula by Hopkins et al. 2005 (see Volonteri, Salvaterra \& Haardt 2006 for details).

In this scenario there is a certain freedom in the 
choice of the seed masses, in the accretion prescription, and in the
MBHB coalescence efficiency. 

In the VHM model, seed MBHs form with masses 
$m_{\rm seed}\sim$ few$\times10^2\msun$, in 
halos collapsing at $z=20$ from rare 3.5-$\sigma$ peaks of the primordial 
density field (Madau \& Rees 2001), and are thought to be the end--product of the first 
generation of stars. 

A different class of models assumes that MBH seeds form already massive. 
In the KBD model, seed MBHs form from the low angular momentum tail of material in halos 
with efficient molecular hydrogen gas cooling. MBHs with mass
\beq 
m_{\rm seed}\simeq5\times10^4\msun\left(\frac{M_H}{10^7\msun}\right)\left(\frac{1+z}{18}\right)^{3/2}\left(\frac{\lambda}{0.04}\right)^{3/2}
\label{KBD:mbh}
\eeq
form in in DM halos with mass  
\beq
M_H \gsim 10^7\msun \left(\frac{1+z}{18}\right)^{-3/2}\left(\frac{\lambda}{0.04}\right)^{-3/2}.
\eeq
We have fixed the free parameters in Eq. \ref{KBD:mbh} by requiring an acceptable match
with the luminosity function (LF) of quasars at $z<6$. We note that, by requiring that the model reproduces the LF, the number of MBH seeds is very much reduced with respect to Koushiappas \& Zentner (2006), where most of the black hole growth was due to black hole mergers. 

Here $\lambda$ is the so called spin parameter, which is a measure of the angular 
momentum of a dark matter halo $\lambda \equiv J |E|^{1/2}/G M_H^{5/2}$, where 
$J$, $E$ and $M_h$ are the  total angular momentum, energy and mass of the halo.  
The angular momentum of galaxies is believed to have been acquired by tidal 
torques due to interactions with neighboring halos. The distribution of spin 
parameters found in numerical simulations is well fit by a lognormal 
distribution in $\lambda_{\rm spin}$, with mean  $\bar \lambda_{\rm spin}=0.04$ 
and standard deviation $\sigma_\lambda=0.5$ (Bullock et al. 2001, van den 
Bosch et al. 2002). We have assumed that the MBH formation process proceeds 
until $z\approx15$.

In the BVR model, black hole  seeds form in halos  subject to runaway gravitational instabilities. 
Gravitational instabilities are likely the most effective process for removing angular momentum.
BVR have suggested that  gas-rich halos with efficient cooling and low angular momentum 
(i.e., low spin parameter) are prone to global dynamical instabilities, 
the so-called ``bars within bars" mechanism (Shlosman, Frank \& Begelman 1989).  
In metal--free halos with virial temperatures $T_{\rm vir} \gsim 10^4$K, hydrogen atomic 
line emission can cool the gas down to $\sim 8000$ K. In smaller 
halos, provided that molecular hydrogen cooling is efficient,  gas can cool 
well below the virial temperature. The amount of material participating in 
the ``bars within bars" instability, however, is much smaller in such mini-halos, leading
to the accumulation in the proto-galaxy centre of only a few  tens of solar masses. 
We assumed here, as in BVR, that MBH seed formation is efficient 
only in metal free halos with virial temperatures $T_{\rm vir} \gsim 10^4$K. 
The ``bars within bars"  process produces in the center of the halo  a ``quasistar" 
(QSS) with a very low specific entropy.  When the QSS  core  collapses, it leads 
to a seed black hole  of a few tens of solar masses.  Accretion from the QSS 
envelope surrounding the collapsed core can however build up a substantial 
black hole mass very rapidly  until it reaches a mass of the order of the "quasistar"  
itself, $M_{\rm QSS}\simeq 10^4-10^5 \msun$.   The black hole accretion rate adjusts 
so that the feedback energy flux equals the Eddington limit for the quasistar 
mass; thus, the black hole grows at a super-Eddington rate as long as 
$M_{QSS} > M_{\rm BH}$ $M_{\rm BH} (t) \sim  4 \times 10^5  (t/10^7 \ {\rm yr})^2  
M_\odot$ i.e., $M_{\rm BH} \propto t^2$.  

In metal rich halos star formation becomes efficient, and depletes the gas 
inflow before the conditions for QSS (and MBH) formation are reached. BVR 
envisage that the process of MBH formation stops when gas is sufficiently 
metal enriched. Given the uncertainties in the efficiency of spreading metals, 
we consider here two scenarios, one in which star formation exerts a high 
level of feedback and ensures a rapid metal enrichment (BVRhf) and one in 
which feedback is milder and halos remain metal free for longer (BVRlf).
In the former case MBH formation ceases at $z\approx 18$, in the latter at $z\approx 15$.
The BVRhf model appears to produce barely enough MBHs to reproduce the 
observational constraints (ubiquity of MBHs in the local Universe, luminosity function of quasars). 
We consider it a very strong lower limit to the number of seeds that need to be formed 
in order to fit the observational constraints. 

\begin{figure}
\centerline{\psfig{file=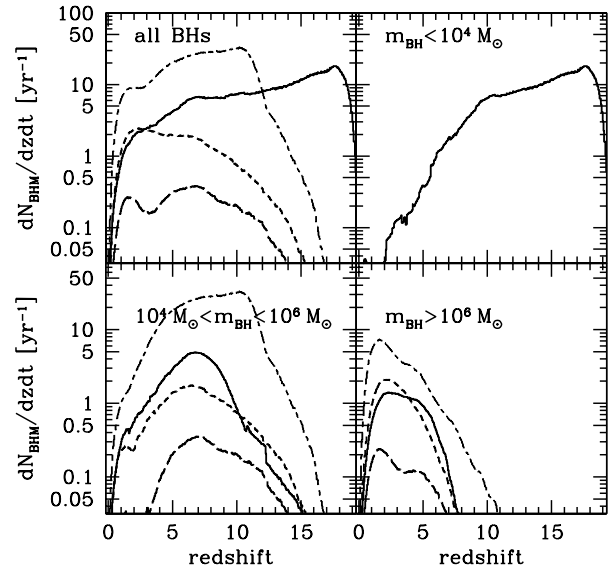,width=84.0mm}}
\caption{Number of MBHB coalescences per observed year at $z=0$, per unit redshift, 
in different $m_{\rm BH}=m_1+m_2$ mass intervals. {\it Solid lines}: VHM model;
{\it short--long dashed lines}: KBD model; {\it short--dashed lines}: 
BVRlf model; {\it long--dashed lines}: BVRhf model.}
\label{mrate}
\end{figure}

Figure \ref{mrate} shows the number of MBH binary coalescences per unit 
redshift per unit {\it observed} year, $dN/dzdt$, predicted by the five models we tested. 
Each panel shows the rates for 
different $m_{\rm BH}=m_1+m_2$ mass intervals. The total coalescence rate
spans almost two orders of magnitude
ranging from $\sim 3$ yr$^{-1}$ (BVRhf) to $\sim 250$ yr$^{-1}$ (KBD).
As a general trend, coalescences of more massive MBHBs peak at lower 
redshifts (for all the models the coalescence peak in the case 
$m_{\rm BH}>10^6\msun$ is at $z\sim2$). 
Note that there are no merging MBHBs with $m_{\rm BH}<10^4\msun$ 
in the KBD and BVR models.

\subsection{MBHB dynamics}
During a galactic merger, the central MBHs initially share their fate with the host galaxy. 
The merging is driven by dynamical friction, which has been shown to efficiently 
merge the galaxies and drive the MBHs in the central regions of the newly formed 
galaxy when the mass ratio of the satellite halo to the main halo is sufficiently 
large, that is when (total) mass ratio of the progenitor halos is, $P=M_s/M \gsim 0.1$ 
(Kazantzidis et al. 2005).  

The efficiency of dynamical friction decays 
when the MBHs get close and form a binary.  In gas-poor systems, the subsequent 
evolution of the binary may be largely determined by three-body interactions 
with background stars (Begelman, Blandford \& Rees 1980), leading to a long coalescence timescale.  
In gas rich high redshift halos, the orbital evolution of the central SMBH is 
likely dominated by dynamical friction against the surrounding gaseous medium. 
The available simulations (Escala et al. 2004; Dotti et al. 2006; Mayer et al. 2006) 
show that the binary can shrink to about parsec or slightly subparsec scale by 
dynamical friction against the gas, depending on the gas thermodynamics. We 
have assumed here that, if a hard MBH binary is surrounded by
an accretion disc, it coalesces  instantaneously owing to interaction with the gas disc. If 
instead there is no gas readily available, the binary will be losing orbital 
energy to the stars, using the scheme described in Volonteri, Madau \& Haardt 
(2003) and in Volonteri \& Rees (2006).

\section{Gravitational wave signals}
Full discussion of the GW signal produced by an inspiraling 
MBHB can be found in Sesana \etal 2005, along with all the relevant references.
Here we just summarise the basic equations.

\subsection{Characteristic strain}
Consider a MBHB at (comoving) distance $r(z)$. The
strain amplitude (sky and polarization averaged) at the rest-frame
frequency $f_r$ is (e.g., Thorne 1987)
\begin{equation}
h\,=\,\frac{8\pi^{2/3}}{10^{1/2}}\,\frac{G^{5/3}{\mathcal M}^{5/3}}{c^4r(z)}\,f_r^{2/3}, 
\label{eqstrain}
\end{equation}
where ${\mathcal M}=m_1^{3/5}m_2^{3/5}/(m_1+m_2)^{1/5}$ is the ``chirp mass'' 
of the binary and all the other symbols have their standard meaning.
The characteristic strain is defined as 
\begin{equation}
h_c=h\sqrt{n} \simeq \frac{1}{3^{1/2}\pi^{2/3}}\,\frac{G^{5/6}
{\mathcal M}^{5/6}}{c^{3/2} r(z)}\,f_r^{-1/6},
\label{eq1h_c}
\end{equation}
where $\sqrt{n}$ is the number of cycles spent in a frequency interval 
$\Delta f \simeq f$. Equation \ref{eq1h_c} is valid only if the
typical source shifting time (the time spent in a given 
$\Delta f \simeq f$ bin) is shorter than the instrumental operation time.
This is almost always the case, as {\it LISA} would be sensitive
to the signal emitted in the last $1-3$ years before the MBHB coalescence, 
and the operation time is expected to be $\gsim 3$ yrs.

\subsection{Resolved events}
An inspiraling binary is then detected if the signal-to-noise ratio ($S/N$)
{\it integrated over the observation} is larger than the assumed
threshold for detection. 
The integrated $S/N$ is given by
(e.g., Flanagan \& Hughes 1998)
\begin{equation}
S/N_{\Delta f}=\sqrt{ \int_{f}^{f+\Delta f} d\ln f' \, \left[
\frac{h_c(f'_r)}{h_{\rm rms}(f')} \right]^2}. 
\label{eqSN}
\end{equation}
Here, $f=f_r/(1+z)$ is the (observed) frequency emitted at time $t=0$ of the observation, 
and $\Delta f$ is the (observed) frequency
shift during the observational time $\tau$.
Finally, $h_{\rm rms}$ is the effective rms noise of the instrument.
The total {\it LISA} $h_{\rm rms}$ noise is obtained by adding 
in quadrature the instrumental rms noise (given by e.g. the Larson's 
online sensitivity curve generator http://www.srl.caltech.edu/$\sim$shane/sensitivity)
and the confusion noise from  unresolved galactic (Nelemans \etal 2001) and
extragalactic (Farmer \& Phinney 2003) WD--WD binaries. Given the uncertainties
on the very-low frequency {\it LISA} sensitivity, we adopt a pessimistic cut
at $10^{-4}$ Hz. We will discuss later the impact of changing the low-frequency 
cut--off of the sensitivity curve.\\
Given a coalescence rate $R$, and the source frequency shift rate
$\dot f$, we can derive the number of {\it individual} binaries resolved with $S/N>s$, 
i.e., (Sesana \etal 2005):
\begin{equation}
N_{\tau}(>s)=R\int_{f_{\rm min}}^{f_{\rm ISCO}}\frac{df}{\dot
f}H_s(\Delta f)
\label{eqbinSN}
\end{equation}
where 
\begin{equation}
H_s(\Delta f)=\cases{1,\,\,\,\, S/N_{\Delta f}\geq s \cr 0,\,\,\,
 S/N_{\Delta f}< s}
\end{equation}
In equation~\ref{eqbinSN}, $f_{\rm ISCO}$ is the observed frequency
emitted at the Keplerian innermost stable circular orbit (ISCO), and  
$f_{\rm min} \ll f_{\rm ISCO}$ is the minimum observed frequency 
of the spiral--in phase.

\section{MBH formation models and the GW signal}

In this section we discuss the characteristics of the GW signal detectable by {\it LISA} as 
predicted by the different models of MBH formation and evolution discussed in section 2. 
All the results shown here assume a {\it LISA} operation time of 3 years, a 
cut-off at $10^{-4}$ Hz in the instrumental sensitivity and a detection 
integrated threshold of $S/N=5$ (eq.~\ref{eqSN}).
\begin{figure}
\centerline{\psfig{file=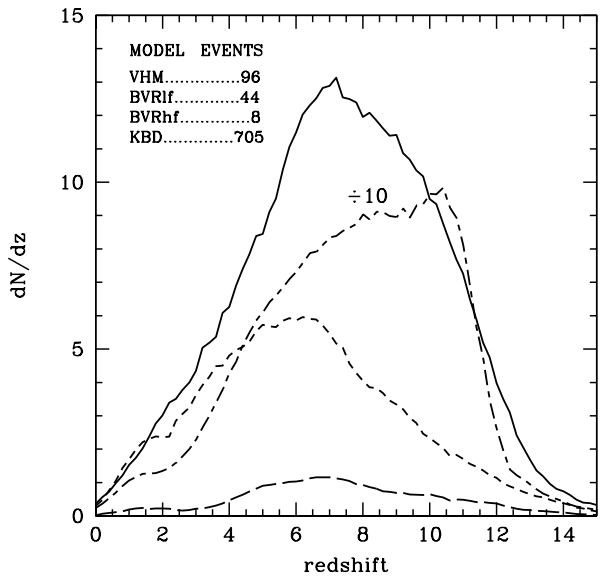,width=84.0mm}}
\caption{Redshift distribution of MBHBs resolved with 
$S/N>5$ by {\it LISA} in a 3-year mission. Line style as in Fig. \ref{mrate}.
The number of events predicted by KBD model
({\it long--short dashed curve}) is divided by a factor of $10$.
The top-left corner label lists the total number of expected detections.
}
\label{zdist}
\end{figure}

\subsection{Event number counts}
Figure~\ref{zdist} shows the redshift distribution of {\it LISA} 
MBHB detections. There are substantial differences between the 
different models. The KBD model results in a number of 
events ($\simeq 700$) that is more than an order of magnitude 
higher than that predicted by other models, 
with a skewed distribution peaked at sensibly high redshift, $z \gsim 10$. 
It is interesting to compare the {\it number of detections} with the {\it total number 
of binary coalescences}  predicted by the different formation models.
The KBD model produces $\simeq 750$ coalescences, the VHM model $\simeq 250$, 
and the two BVR models just few tens. A difference of a factor $\simeq 3$  
between the KBD and the VHM models in the total number of coalescences,
results in a difference of a  factor of $\simeq 10$ in the {\it LISA} detections, 
due to the different mass of the seed black holes. 
Almost all the KBD coalescences involve massive binaries ($m_1 \gsim 10^4 \msun$), 
which are observable by  {\it LISA}. The  KBD  and BVR models differ for 
the sheer number of MBHs. The halo mass threshold in the KBD model is well below 
(about 3 orders of magnitude) the BVR one, the latter requiring halos with virial 
temperature above $10^4$K. In a broader context,  results pertaining to the 
KBD model describe the behaviour  of families of models where efficient 
MBH formation can happen also in mini-halos where the source of cooling is molecular hydrogen. 

It is difficult, on the basis of the redshift distributions of detected binaries only, 
to discriminate between heavy and light MBH seed scenarios.  
Although the VHM and BVRlf models predict a different number of observable sources,
the uncertainties in the models are so high, that a difference 
of a factor of two (96 for the VHM model, 44 for the BVRlf model) cannot be 
considered a safe discriminant. Moreover the redshift distributions are quite
similar, peaked at $z\simeq6-7$ and without any particular feature in the shape. 
Arguably, the initial mass distribution of seeds is the main variable
influencing the number of {\it LISA} events.  
However, a factor of two difference in the number counts, in fact, can be ascribed to 
uncertainties in other aspects of the MBH evolution, both dynamical and related to 
the accretion history. Sesana et al. (2005) showed how uncertainties in 
the dynamical evolution timescales reflects on the detectable events. 
Also the accretion history plays a role which may not be marginal.  
For instance, if MBHs grow fast (accreting around the Eddington rate or higher), the number of coalescences
that can be detected by LISA is higher than, say, a case in which the accretion rate is sub-Eddington 
for most sources. 

\begin{figure}
\centerline{\psfig{file=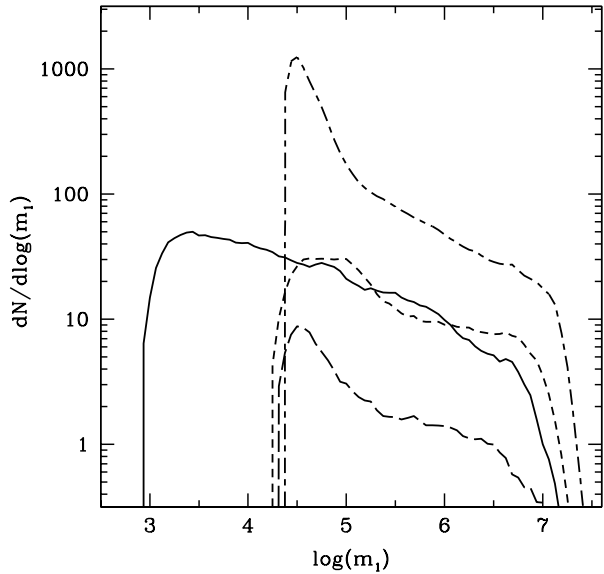,width=84.0mm}}
\caption{Mass function of the more massive member 
of MBHBs resolved with $S/N>5$ by {\it LISA} in a 3-year mission. 
Line style as in figure \ref{mrate}. All curves are normalized such as the integral 
in $d\log(m_1)$ gives the number of detected events.}
\label{m1dist}
\end{figure}
\begin{figure}
\centerline{\psfig{file=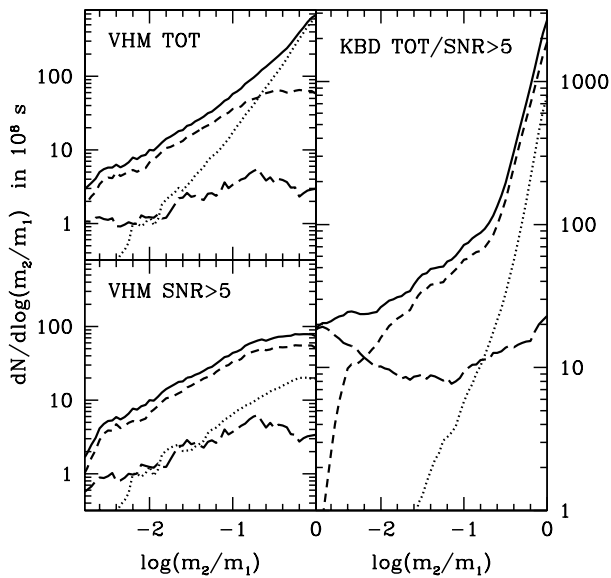,width=84.0mm}}
\caption{Mass ratio distribution of  MBHBs.  {\it long--dashed curve:} $0<z<3$; 
{\it short--dashed curve:} $3<z<10$; {\it dotted curve:} $z>10$; {\it solid curve:} 
all redshifts. {\it Left panels:} VHM model, all coalescences ({\it upper panel}), and 
coalescences detectable by {\it LISA} with $S/N>5$ in a 3-year mission ({\it lower panel}).  
Most high-redshift events with mass ratios of order unity involve light 
binaries which cannot be observed by {\it LISA}. 
{\it Right panel:} KBD model; almost all the coalescences can be observed with a $S/N>5$.}
\label{mrdist}
\end{figure}

\subsection{Black hole masses and mass ratio distributions}
In Sesana \etal 2005 we showed that {\it LISA} will be sensitive to 
binaries with masses $\lsim10^3\msun$ up to redshift ten. Hence the 
discrimination between heavy and light MBH seed scenarios should be 
easy on the basis of the mass function of detected binaries.
This is shown in figure~\ref{m1dist}. 
As expected, in the VHM model, the mass distribution extends to masses $\lsim10^3\msun$,
giving a clear and unambiguous signature of a light seed scenario. 
VHM  predict that  many detections (about 50$\%$) involve 
low mass binaries ($m_{\rm BH}<10^{4}\msun$) at high redshift ($z>8$). These sources
are observable during the inspiral phase, but their $f_{\rm ISCO}$ is too high for 
{\it LISA} detection (see Sesana \etal 2005, figure 2). 
Heavy seed scenarios predict instead that the GW emission at $f_{\rm ISCO}$, and the subsequent
plunge are always observable for all binaries. 

In figure \ref{mrdist} we show the mass ratio distribution of the resolved events
for the VHM and the KBD models. Model KBD (as well as BVRlf 
and BVRhf, not shown here) predict a monotonically increasing distribution, 
the majority of detections having a mass ratio $q=M_2/M_1 \leqslant 1$.
A large fraction of observable 
coalescences, in fact, involve MBHBs at $z>10$, when MBHs had no time to accrete 
much mass yet. As most seeds form with similar mass ($\simeq 10^4 \msun$, see KBD; 
$\simeq 10^2 \msun$, see VHM), mergers at early times involve MBHBs with $q \simeq 1$. 
In massive seeds scenarios, almost all coalescences are observable, and the mass ratio 
distribution is dominated by $z>10$ mergers between seeds ($q \simeq 1$).
In scenarios based on Population III remnants, $z>10$ mergers involve MBHs with 
mass below the {\it LISA} threshold. The detectable events happen at later 
times, when MBHs have already experienced a great deal of mass growth. VHM 
models therefore produce a mass ratio distribution which is flat or features a broad 
peak at $q \simeq 0.1-0.2$, depending on the details of the  accretion prescription.  
This is due to the fact that both the probability of halo mergers (because of the 
steep DM halo mass function) and the dynamical friction timescale increase with 
decreasing halo mass ratio. Hence, equal mass mergers that lead to efficient binary formation within short timescales (i.e., shorter than the Hubble time) are rare,  while in more common unequal mass mergers it takes longer than an Hubble time to drag  the satellite hole to the center.

All the results shown above assume 3 yrs observation and a cut-off
in the {\it LISA} sensitivity curve at $10^{-4}$ Hz. 
Even in the pessimistic case of a 1 yr mission lifetime, however, 
with a sensitivity curve cut at $10^{-4}$ Hz and assuming the BVRhf seed model, 
{\it LISA} is expected to observe at least two or three MBHB merging events. 
We stress once again that the BVRhf model provides a strong lower limit to the 
number and redshift distribution of forming seeds, on the basis of current 
observational constraints. 

\begin{figure}
\centerline{\psfig{file=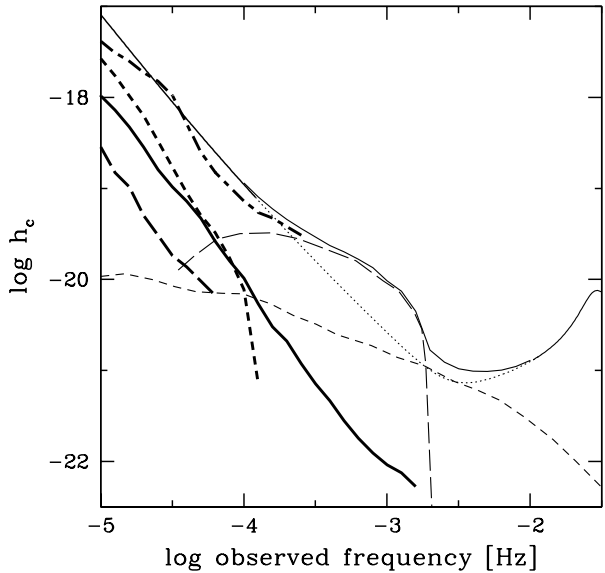,width=84.0mm}}
\caption{Predicted confusion noises assuming a {\it LISA} operating time 
of one year. {\it Thin lines}: {\it LISA} rms confusion noise ({\it solid line}), as
the quadratic sum of the {\it LISA} instrumental single--arm Michelson
noise ({\it dotted line}, from 
http://www.srl.caltech.edu/$\sim$shane/sensitivity), and the confusion noise from  
unresolved galactic (Nelemans \etal 2001, {\it long-dashed line}), and
extragalactic (Farmer \& Phinney 2003, {\it short-dashed line})
WD--WD binaries. {\it Thick lines}: predicted confusion noises 
for the different MBHB models we tested. 
Line style as in in figure~\ref{mrate}.}
\label{confnoise}
\end{figure}

\subsection{Confusion noise}
If the number of merging sources is so large that there are, 
on average, at least eight sources above threshold per frequency resolution bin, 
then the total signal will be observed as a confusion noise (Cornish 2003).
A detectable confusion noise of cosmic origin would provide much information
on the emitting population, but, on the other hand, would be
added (in quadrature) to the instrumental noise, reducing 
the interferometer capability of detecting individual sources.
Assuming a mission lifetime of three yrs, the predicted confusion 
noise (see Sesana \etal 2005 for details), varies by an order of magnitude for 
different models, but lies,  for all models, below the {\it LISA} sensitivity
curve. In the pessimistic view of a one year mission, the confusion noise is enhanced
roughly by a factor of three. As shown in figure~\ref{confnoise},
the confusion noise predicted by the KBD model is expected to be 
comparable to the rms noise at frequencies $\lsim3\times10^{-4}$ Hz.
If the sensitivity curve cuts--off at $10^{-5}$ Hz;
the quadrature addition of such a noise would result in a slight 
decrease of the total {\it LISA} sensitivity in the frequency range 
$3\times 10^{-5}-3\times 10^{-4}$ Hz. 

\section{Summary and conclusions}
Using dedicated Montecarlo simulations of the hierarchical assembly 
of DM halos along the cosmic history, we have 
computed the expected gravitational wave signal from the evolving 
population of massive black hole binaries. 
The imprint of black hole mergers and coalescences on the {\it LISA} 
data stream depends on the specific assumptions 
regarding MBH formation, and on the recipes employed for 
the hole mass growth via merger and gas accretion. 

We have considered two main frameworks for MBH formation, namely, one where 
seeds are light ($\simeq 10^2 \msun$),  and one where seeds are heavy 
($\gsim 10^4 \msun$).  In the former, MBH seeds form at $z\simeq 20$ with 
masses of few hundreds solar masses, and are though to be  the endpoint 
of the evolution of metal--free massive stars (VHM). 
In the heavy seeds scenarios, MBHs form in the centers of high-redshift gas-rich 
halos where angular momentum losses are efficient. KBD explore a model where 
angular momentum is shed via turbulent viscosity, in all halos with efficient 
molecular hydrogen cooling. This seed formation scenario is very efficient, 
and predicts that seeds are widespread, forming in halos as small as a few 
$10^5 \msun$, provided that the total angular momentum of the halo is small
enough. BVR explore a different scenario where angular momentum is transported 
via runaway gravitational instabilities ("bars-within-bars"). BVR envisage 
that the process would be more effective in halos with efficient atomic 
cooling, that is with virial temperature  $T_{\rm vir} \gsim 10^4$K, and 
mass $M_h\gsim 10^8 \msun$. MBH seeds are therefore much rarer in the BVR 
model. BVR envisage that the process of MBH formation stops when gas is 
sufficiently metal enriched. Given the uncertainties in the efficiency in 
spreading metals, we consider here two scenarios, one in which star formation 
exerts a high level of feedback and ensures a rapid metal enrichment (BVRhf), 
one in which feedback is milder and halos remain metal free for longer 
(BVRlf). In the former case MBH formation ceases at $z\approx 18$, in 
the latter at $z\approx 15$.

We have shown that, in all considered models, MBHB coalescences do not 
produce a stochastic GW background, but rather, a set of individual 
resolved events. A large fraction (depending on models) of coalescences 
will be directly observable by {\it LISA}, and on the basis of the detection 
rate, constraints can be put on the MBH formation process. 
Detection of several hundreds events in 3 years will be the sign of 
efficient formation of heavy MBH seeds in a large fraction of high redshift halos (KBD).

At the other extreme, a low event rate, about  few tens in 3 years,  
is peculiar in scenarios where either the seeds are light, and many 
coalescences do not fall into the {\it LISA} band, or seeds are massive, 
but rare, as envisioned by, e.g., BVR (see also Lodato \& Natarajan). 
In this case a decisive diagnostic is provided by the mass distribution 
of detected events. In the light seed scenario, the mass distribution
of observed binaries extend to $\sim 10^3\msun$, while there are no
sources with mass below $10^4\msun$ in the high seed scenario.  
Finally, we have shown that a further, helpful diagnostic of MBH models 
lies in the distribution of the mass ratios in binary 
coalescences. While heavy seed models predicted that most of the 
detected events involve equal mass binaries, 
in the case of light seeds, mass ratios are equally distributed in the
mass ratio range $0.1-1$.  

Should the early black hole population be dominated by massive systems
(e.g., KBD), the GW signal can be accompanied by an electromagnetic counterpart (EM, Miloslavljevic \& Phinney 2005, Dotti et al. 2006, Kocsis et al. 2006), in principle detectable by future high-sensitivity X--ray telescopes (e.g., XEUS\footnote{www.rssd.esa.int/index.php?project=XEUS}). 
The identification of an EM counterpart would have crucial implications for cosmology. {\it LISA} sources could then be used as ``standard sirens'' (Schutz 1986) to estimate fundamental cosmological parameters (Schutz 2002; Holz \& Hughes 2005), providing constraints complementary to cosmic microwave background experiments, and supernovae search experiments. 

We note, however, that from the astrophysical point of view, even in absence of EM counterparts, by adopting the standard Wilkinson Microwave Anisotropy Probe cosmological parameters (Spergel et al. 2003) the redshift of the merging MBHs will be known with the same precision as most sources observed in electromagnetic bands. 

In conclusion, concerning the detection of low frequency 
gravitational waves, massive black hole binaries are certainly one of 
the major target for a mission as {\it LISA}. On the astrophysical ground, 
{\it LISA} will be a unique probe of the formation, 
accretion and merger of MBHs along the {\it entire} cosmic history of galactic structures.

\vspace{+0.5cm}
We would like to thank A. Vecchio for helpful discussions and comments on our 
manuscript. This research was supported in part by the National 
Science Foundation under Grant No. PHY99-07949. 

{}

\end{document}